\documentclass[twocolumn,showpacs,preprintnumbers]{revtex4}
\usepackage{amssymb}

\usepackage[dvips]{graphicx}

\input{tcilatex}

\begin{document}

\title{Comment on \emph{Hysteresis, Switching, and Negative Differential Resistance
in Molecular Junctions: a Polaron Model}, by M. Galperin, M.A. Ratner, and
A. Nitzan, Nano Lett. \textbf{5}, 125 (2005)}
\author{A. S. Alexandrov$^{1}$ and A.M. Bratkovsky$^{2}$}

\begin{abstract}
It is shown that the ``hysteresis'' in a polaron model of electron transport
through the molecule found by M.~Galperin \textit{et al}. [Nano Lett. 
\textbf{5}, 125 (2005)] is an artefact of their ``mean-field''
approximation. The reason is trivial:\ after illegitimate replacement $\hat{n%
}^{2}=\hat{n}n_{0},$ where $n_{0}=\left\langle c_{0}^{\dagger
}c_{0}\right\rangle \leqslant 1$\thinspace\ the average molecular level
occupation Galperin \textit{et al} obtained non-physical dependence of a
nondegnerate molecular energy level on the non-integer mean occupation
number $n_{0}$\ (i.e. the electron self-interaction) and the resulting
non-linearity of current. The theory of correlated polaronic transport
through molecular quantum dots (MQDs) that we proposed earlier [Phys. Rev. B%
\textbf{67}, 235312 (2003)] proved that there is no hysteresis or switching
in current-voltage characteristics of non-degenerate, $d=1$, or double
degenerate, $d=2$, molecular bridges, contrary to the mean-field result.
Switching could only appear in multiply degenerate MQDs with $d>2$ due to
electron correlations. Most of the molecular quantum dots are in the regime
of weak coupling to the electrodes addressed in our formalism.
\end{abstract}

\affiliation{$^{1}$Department of Physics, Loughborough University, Loughborough LE11 3TU,
United Kingdom\\
$^{2}$Hewlett-Packard Laboratories, 1501 Page Mill Road, MS 1123, Palo Alto,
California 94304\\
}
\maketitle

Although the correlated electron transport through mesoscopic systems with
repulsive electron-electron interactions received considerable attention in
the past, and continues to be the focus of intensive studies \cite{molnano},
much less has been known about a role of electron-phonon correlations in
MQD. Recently we have proposed a negative$-U$ Hubbard model of a $d$-fold
degenerate quantum dot \cite{alebrawil} and a polaron model of resonant
tunneling through a molecule with $d-$degenerate level\cite{alebra}. We
found that the \emph{attractive} electron correlations caused by any
interaction within the molecule could provide a molecular switching effect
where the current-voltage (I-V) characteristic has two branches with high
and low current at the same voltage. This prediction has been confirmed by
our theory of the \textit{correlated} transport through degenerate MQDs with
a full account of both the Coulomb repulsion and realistic electron-phonon
(e-ph) interactions \cite{alebra}. We have shown that while the phonon
side-bands significantly modify switching in comparison with the negative-$U$
Hubbard model (appearance of phonon ladder on the I-V curve), switching is
robust. It shows up when the effective interaction of polarons is attractive
and the state of the dot is multiply degenerate, $d>2$, while there is \emph{%
no switching} in a non-degenerate ($d=1)$ or a double degenerate ($d=2)$ MQD.

Surprisingly, later on Galperin \textit{et al}. \cite{rat} neglected these
results, claiming that even a non-degenerate electronic level coupled to a
single vibrational mode provides an I-V curve with the hysteresis,
switching, and negative differential resistance. Here we show that these
findings are artefacts of their mean-field approximation that neglects the
Fermi-Dirac statistics of electrons.

First, we will illustrate the failure of the mean-field approximation of
Ref.~\cite{rat} on a simplest model of a single atomic level coupled with a
single one-dimensional oscillator using the first quantization
representation for its displacement $x$, 
\begin{equation}
H=\varepsilon _{0}\hat{n}+fx\hat{n}-{\frac{1}{{2M}}}{\frac{\partial ^{2}}{{%
\partial x^{2}}}}+{\frac{kx^{2}}{{2}}}.
\end{equation}
Here $M$ and $k$ are the oscillator mass and the spring constant, $f$ is the
interaction force, and $\hbar =c=k_{B}=1$. This Hamiltonian is readily
diagonalized with the \emph{exact} displacement transformation of the
vibration coordinate $x$, 
\begin{equation}
x=y-\hat{n}f/k,
\end{equation}
to the transformed Hamiltonian without electron-phonon coupling, 
\begin{eqnarray}
\tilde{H} &=&\varepsilon \hat{n}-{\frac{1}{{2M}}}{\frac{\partial ^{2}}{{%
\partial y^{2}}}}+{\frac{ky^{2}}{{2}}},  \label{eq:newH} \\
\varepsilon  &=&\varepsilon _{0}-E_{p},  \label{eq:newLev}
\end{eqnarray}
where we used $\hat{n}^{2}=\hat{n}$ because of the Fermi-Dirac statistics.
It describes a small polaron at the atomic level $\varepsilon _{0}$ shifted
down by the polaron level shift $E_{p}=f^{2}/2k$, and entirely decoupled
from ion vibrations. The ion vibrates near a new equilibrium position,
shifted by $f/k$, with the ``old'' frequency $(k/M)^{1/2}$. As a result of
the local ion deformation, the total energy of the whole system decreases by 
$E_{p}$ since a decrease of the electron energy by $-2E_{p}$ overruns an
increase of the deformation energy $E_{p}$. It becomes clear that the major
error of the mean-field approximation of Ref.~\cite{rat} originates in
illegitimate replacement of the square of the occupation number operator $%
\hat{n}=c_{0}^{\dagger }c_{0}$ by its ``mean-field'' expression $\hat{n}^{2}=%
\hat{n}n_{0}$ with the average population of a single molecular level, $n_{0}
$, in disagreement with the exact identity, $\hat{n}^{2}=\hat{n}$. This
leads to a spurious self-interaction of a single polaron with itself [$%
\varepsilon =\varepsilon _{0}-n_{0}E_{p}$ instead of Eq.(\ref{eq:newLev})],
and a resulting non-existent nonlinearity in the rate equation.

The correct procedure should be as follows, see Ref.\cite{alebra}. The
appropriate molecular Hamiltonian includes the Coulomb repulsion, $U^{C}$,
and the electron-vibron interaction as \cite{alebra} 
\begin{eqnarray}
&&H =\sum_{_{\mu }}\varepsilon _{_{\mu }}\hat{n}_{_{\mu }}+\frac{1}{2}%
\sum_{_{\mu }\neq \mu ^{\prime }}U_{\mu \mu ^{\prime }}^{C}\hat{n}_{_{\mu }}%
\hat{n}_{\mu ^{\prime }}  \nonumber \\
&&+\sum_{\mu ,q}\hat{n}_{_{\mu }}\omega _{q}(\gamma _{\mu
q}d_{q}+H.c.)+\sum_{q}\omega _{q}(d_{q}^{\dagger }d_{q}+1/2).
\end{eqnarray}
Here $d_{q}$ annihilates phonons, $\omega _{q}$ is the phonon (vibron)
frequency, and $\gamma _{\mu q}$ are the e-ph coupling constant ($q$
enumerates the vibron modes). This Hamiltonian conserves the occupation
numbers of molecular states $\hat{n}_{_{\mu }}$.

One can apply the canonical unitary transformation $e^{S}$, with $%
S=-\sum_{q,\mu }\hat{n}_{\mu }\left( \gamma _{\mu q}d_{q}-H.c.\right) $
integrating phonons out. The electron and phonon operators are transformed
as 
\begin{equation}
\tilde{c}_{\mu }=c_{\mu }X_{\mu },\qquad X_{\mu }=\exp \left( \sum_{q}\gamma
_{\mu q}d_{q}-H.c.\right)
\end{equation}
and 
\begin{equation}
\tilde{d}_{q}=d_{q}-\sum_{\mu }\hat{n}_{\mu }\gamma _{\mu q}^{\ast },
\end{equation}
respectively. This Lang-Firsov transformation shifts ions to new equilibrium
positions with no effect on the phonon frequencies. The diagonalization is 
\emph{exact}: 
\begin{equation}
\tilde{H}=\sum_{i}\tilde{\varepsilon}_{_{\mu }}\hat{n}_{\mu }+\sum_{q}\omega
_{q}(d_{q}^{\dagger }d_{q}+1/2)+{\frac{1}{{2}}}\sum_{\mu \neq \mu ^{\prime
}}U_{\mu \mu ^{\prime }}\hat{n}_{\mu }\hat{n}_{\mu ^{\prime }},
\end{equation}
where 
\begin{equation}
U_{\mu \mu ^{\prime }}\equiv U_{\mu \mu ^{\prime }}^{C}-2\sum_{q}\gamma
_{\mu q}^{\ast }\gamma _{\mu ^{\prime }q}\omega _{q},  \label{eq:Umm1}
\end{equation}
is the renormalized interaction of polarons comprising their interaction via
molecular deformations (vibrons) and the original Coulomb repulsion, $U_{\mu
\mu ^{\prime }}^{C}$. The molecular energy levels are shifted by the polaron
level-shift due to a deformation created by the polaron, 
\begin{equation}
\tilde{\varepsilon}_{_{\mu }}=\varepsilon _{_{\mu }}{-}\sum_{q}|\gamma _{\mu
q}|^{2}\omega _{q}.  \label{eq:eps}
\end{equation}

Applying the same transformation to the retarded Green's function, one
obtains the exact MQD spectral function \cite{alebra} for a $d-\mathrm{fold}$
degenerate MQD (i.e. the density of molecular states, DOS) as 
\begin{eqnarray}
&&\rho (\omega )=\mathcal{Z}d\sum_{r=0}^{d-1}Z_{r}(n)\sum_{l=0}^{\infty
}I_{l}\left( \xi \right)   \nonumber \\
&&\times \biggl[e^{\beta \omega _{0}l/2}\left[ (1-n)\delta (\omega
-rU-l\omega _{0})+n\delta (\omega -rU+l\omega _{0})\right]   \nonumber \\
&&+(1-\delta _{l0})e^{-\beta \omega _{0}l/2}[n\delta (\omega -rU-l\omega
_{0})  \nonumber \\
&&+(1-n)\delta (\omega -rU+l\omega _{0})]\biggr],  \label{eq:rho}
\end{eqnarray}
where 
\begin{equation}
\mathcal{Z}=\exp \left[ -|\gamma |^{2}\coth \frac{\beta \omega _{0}}{2}%
\right] ,
\end{equation}
is the \emph{polaron} \emph{narrowing factor} \cite{mah}, $\xi =|\gamma
|^{2}/\sinh (\beta \omega _{0}/2),$ $I_{l}\left( \xi \right) $ the modified
Bessel function, $\beta =1/T$, and $\delta _{lk}$ the Kroneker symbol. To
simplify our discussion, we assume that the Coulomb integrals do not depend
on the orbital index, i.e. $U_{\mu \mu ^{\prime }}=U$, and consider a
coupling to a single vibrational mode, $\omega _{q}=\omega _{0}$.

The important feature of DOS, Eq.(\ref{eq:rho}), is its nonlinear dependence
on the average electronic population $n=\left\langle c_{\mu }^{\dagger
}c_{\mu }\right\rangle $ which leads to the switching, hysteresis, and other
nonlinear effects in I-V characteristics for $d>2$ \cite{alebra}. It appears
due to \emph{correlations} between \emph{different} electronic states via
the correlation coefficients 
\begin{equation}
Z_{r}(n)=\frac{(d-1)!}{r!(d-1-r)!}n^{r}(1-n)^{d-1-r}.
\end{equation}
There is no nonlinearity if the dot is nondegenerate, $d=1,$ since $%
Z_{0}(n)=1$, contrary to Ref.~\cite{rat}. In this simple case the DOS, Eq.~(%
\ref{eq:rho}), is a \emph{linear} function of the average population that
can be found as a textbook example of an exactly solvable problems \cite{mah}%
. As a result, the rate equation for $n$ \cite{alebra} yields only a single
solution [see Eq.~(43) in Ref.~\cite{alebra}] and no switching in the whole
voltage range.

However, their ``mean-field'' approximation led the authors of Ref.~\cite
{rat} to the opposite conclusion. Indeed, Galperin \textit{et al}.~\cite{rat}
have replaced the occupation number operator $\hat{n}$ in the e-ph
interaction by the average population $n_{0}$ [Eq.~(2) of Ref.~\cite{rat}]
and found the average steady-state vibronic displacement $\langle
d+d^{\dagger }\rangle $ proportional to $n_{0}$ (this is an\emph{\ }explicit%
\emph{\ neglect of} all \emph{quantum fluctuations} on the dot accounted for
in the exact solution \cite{alebra}). Then, replacing the displacement
operator $d+d^{\dagger }$ in the bare Hamiltonian, Eq.~(1), by its average,
Galperin \textit{et al.} obtained a new molecular level, $\tilde{\varepsilon}%
_{0}=\varepsilon _{0}-2\varepsilon _{reorg}n_{0}$ shifted linearly with the
average population of the level. This is in stark disagreement with the
conventional constant polaronic level shift, Eq.~(\ref{eq:newLev},\ref
{eq:eps}) ($\varepsilon _{reorg}$ is $|\gamma |^{2}\omega _{0}$ in our
notations). Their spectral function turned out to be highly nonlinear as a
function of the population, e.g. for the weak-coupling with the leads $\rho
(\omega )=\delta (\omega -\varepsilon _{0}-2\varepsilon _{reorg}n_{0}),$ see
Eq.~(17) in Ref.~\cite{rat}. As a result, the authors of Ref. \cite{rat}
have found multiple solutions for the steady-state population, Eq.~(15) and
Fig.~1, and switching, Fig.~4 of Ref.~\cite{rat}, which actually do not
exist. Taking into account the coupling with the leads ($\Gamma $ in Ref. 
\cite{alebra,rat}) beyond the second order and the coupling between the
molecular and bath phonons does not provide any non-linearity of the
non-degenerate DOS either, because these couplings do not depend on the
electron population.

Note that the the mean-field solution by Galperin \textit{et al}. \cite{rat}
applies at any ratio $\Gamma /\omega _{0},$\ including the limit of interest
to us, $\Gamma \ll \omega _{0}.$ where their transition between the states
with $n_{0}=0$ and $1$ only sharpens, but none of the results change.
Therefore, they do predict a current bistability in the system where it does
not exist at $d=1.$ Results in Ref.~\cite{rat} are plotted for $\Gamma
\gtrsim \omega _{0},$ $\Gamma \approx 0.1-0.3$ eV, which corresponds to
molecular bridges with a resistance of about a few $100$~k$\Omega .$ Such
model `molecules' are rather `metallic' in their conductance and could
hardly show any bistability because carriers do not have time to interact
with vibrons on the molecule. This obvious conclusion for molecules strongly
coupled to the electrodes can be reached in many ways, see e.g. a very
involved derivation in Ref.~\cite{millis05}. The current hysteresis does not
occur in their model, it remains a single-valued function of bias with
superimposed noise. In any case, it certainly has nothing to do with our
result \cite{alebra} that applies to molecular quantum dots ($\Gamma \ll
\omega _{0})$ with $d>2.$

As a matter of fact, most of the molecules are very resistive, so the actual
molecular quantum dots are in the regime we study, see Ref.\cite{mqdexp}.
For example the resistance of fully conjugated three-phenyl ring Tour-Reed
molecules chemically bonded to metallic Au electrodes \cite{reedndr99} is
larger than $1$G$\Omega $. Different from the non-degenerate dot, the rate
equation for a multi-degenerate dot, $d>2$, weakly coupled to the leads has
multiple physical roots in a certain voltage range and a hysteretic behavior
due to \emph{correlations} between different electronic states of MQD \cite
{alebra}. We conclude that Galperin \textit{et al}. \cite{rat} have found a
non-existent hysteresis in a model already solved well before their work in
Ref.~\cite{alebra}.

\end{document}